\documentclass[aip,prl,reprint,graphicx, amsmath]{revtex4-1}
\usepackage{graphicx}
\usepackage{gensymb}

\draft 

\begin{document}


\title{Enhancement of betatron X-rays through asymmetric laser wakefield generated in transverse density gradients} 

\author{J. Ferri}
\email{julien.ferri@polytechnique.edu}
\affiliation{CEA, DAM, DIF, 91297 Arpajon, France}
\affiliation{LOA, ENSTA ParisTech, CNRS, Ecole Polytechnique, Universit\'e Paris-Saclay, 91762 Palaiseau, France}
\affiliation{Department of Physics, Chalmers University of Technology, 41296 Gothenburg, Sweden}
\author{X. Davoine}
\email{xavier.davoine@cea.fr}
\affiliation{CEA, DAM, DIF, 91297 Arpajon, France}

\date{\today}

\begin{abstract}
Laser wakefield acceleration of electrons usually offers an axisymmetry around the laser propagation axis. Thus, the accelerating electrons that are focused on axis often execute small transverse oscillations. In this Article, we propose a simple scheme to break this symmetry, which enhances the transverse wiggling of electrons and boosts the betatron radiation emission. Through 3D particle-in-cell simulations, we show that sending the laser with a small angle of incidence on a transverse plasma density gradient generates an asymmetric wakefield. It first provokes injection and then increases the wiggling of the electrons through the transverse shifting of the wakefield axis which occurs when the laser pulse leaves the gradient. Consequently, we show that the radiated energy per unit of charge can increase by a factor $>20$ when using this scheme, and that the critical energy of the radiation quintuples compared with a reference case without the transverse density gradient.
\end{abstract}


\maketitle 


\section{Introduction}
When propagating in low-density gas jets, ultra-short laser pulses can generate plasma waves characterized by high amplitude electric fields, which, in turn, can be used to accelerate electron bunches to 100's of MeV or GeV energies in millimetric or centimetric distances \cite{taji79, faur04, gedd04, mang04, leem06}. Electron acceleration has been proved to be particularly efficient when the laser duration and waist are matched with the plasma density, in the blowout \cite{lu06a, lu06b} regime. Moreover, when using external injection methods like longitudinal gradient injection \cite{bula98, gedd08}, high-quality and tunable quasi-monoenergetic electron bunches can be generated in reproducible experiments \cite{gons11,hans15}. Additionally, the transverse motion of the electrons during their acceleration generates betatron radiation: a high brilliance, synchrotron-like X-ray source with broad spectrum in the keV to 10's of keV range \cite{esar02}.
In these last years, the main features of this source have been characterized and improved. Micrometric size has been measured \cite{shah06} and duration has been shown to be inferior to 100~fs \cite{taph07} (measures made on the electron beam even suggest a shorter duration of a few femtoseconds \cite{lund11}), whereas divergence smaller than 10 milliradians and photon energies up to several tens of keV were obtained \cite{knei10}, outclassing the first results obtained in 2004 \cite{rous04}. These improvements recently enabled the use of this kind of source for high resolution X-ray phase contrast imaging of insects samples \cite{four11,knei11}.
However, for a wide range of applications, the photon flux shall necessarily be further increased so the betatron source can concurrence the X-ray sources obtained in conventional accelerators \cite{albe14}. For example, medical X-ray phase contrast imaging of human bones requires more energetic photons than those which were obtained in experiments performed on more transparent insects samples \cite{cole15}. 

If we neglect the electron acceleration, the betatron X-ray spectrum is characterized by the distribution function \cite{esar02} $S(\omega/\omega_c) = \int_{\omega/\omega_c}^{\infty}K_{5/3}(x)dx$, where $K_{5/3}$ is a modified Bessel function of the second kind and $\omega_c = E_c/\hbar$, with $\hbar$ the reduced Planck constant and $E_c$ the critical energy given by $E_c[\text{keV}] = 5.24\times10^{-24}\gamma^2n_e[\text{cm}^{-3}]r_{\beta}[\mu\text{m}]$. In this last formula, $\gamma$ is the Lorentz factor of the electrons, $n_e$ is the plasma density and $r_{\beta}$ is the amplitude of the electron transverse motion (oscillation radius). In addition, the number of photons emitted per electron oscillation is $N_X = 4.4 \times 10^{-12}\sqrt{\gamma n_e[\text{cm}^{-3}]}r_{\beta}[\mu\text{m}]$.
Both these values increase with any of the three parameters $\gamma$, $n_e$ or $r_{\beta}$, defining different ways of increasing the X-ray emission. Higher electron energies are naturally attained as higher laser powers become available \cite{kim13, wang13, leem14}. However, in order to develop more accessible high-resolution X-ray sources, it is important to improve the betatron source flux independently of the laser energy growth. Recent work suggests that this shall be achieved by improving the laser beam quality \cite{ferr16}. Another possibility could consist in an increase of the plasma density, but $n_e$ cannot be much modified, as increasing the density leads to a reduction of the dephasing length, and thus of the electron energy. An optimal density between a low value that favors the energy gain and a high value that favors a strong electron wiggling is thus usually found experimentally, or a two-stage scheme has to be adopted \cite{ferr18}. The last solution lies in an increase of $r_{\beta}$. This has been proposed through the resonant interaction of the accelerated beam with the rear of the laser pulse \cite{neme08}, but one has to be quite careful when modeling this phenomenon with PIC codes \cite{lehe14}.

\begin{figure*}[!t!]
\centering
\includegraphics[width=0.95\textwidth]{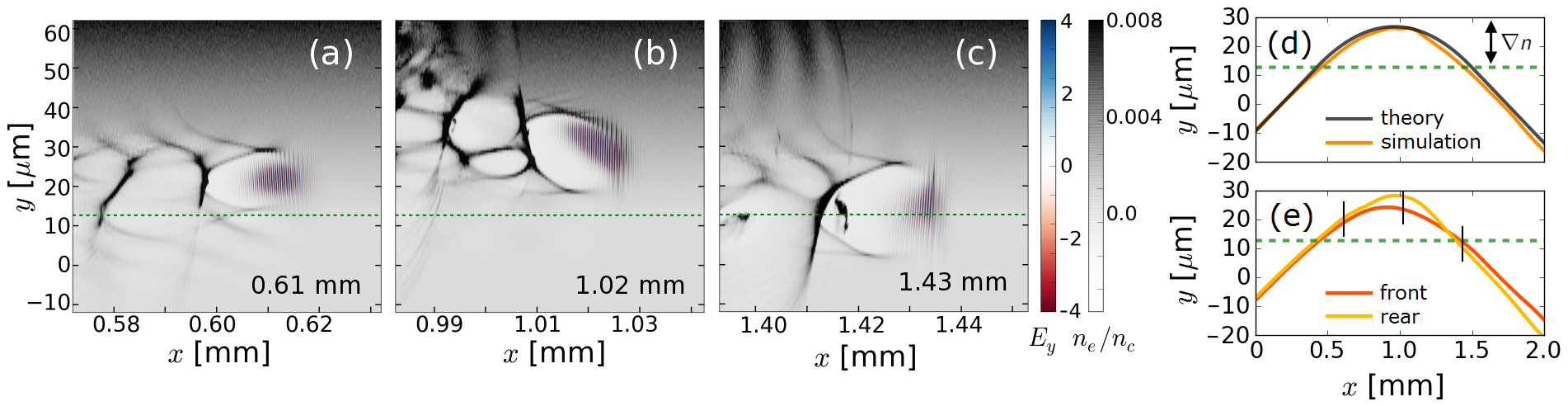}
\centering
\caption{Laser deflection and wakefield asymmetry in the transverse density gradient. (a)-(c) Plasma density (gray), and laser field (in units of $m_ec\omega_0/e$, red-blue) after different propagation distances. (d) Trajectory of the laser beam barycenter in the simulation (orange) and theoretical trajectory of a light ray (black) and (e) trajectories of the front (first $c/\omega_p$ of the pulse, red) and the rear (yellow) of the laser pulse. The bottom of the transverse gradient is added in dashed green line. The initial angle of incidence is $\alpha=3\degree$, and the time steps corresponding to figures (a)-(c) are represented by the vertical black bars in (e).}
\label{champlas}
\end{figure*}

In this Article, we propose a new scheme based on the manipulation of the plasma density profile, in order to increase the number of high-energy photons emitted by the betatron source. Laser wakefield acceleration generally presents an axisymmetry, which leads to the injection of a well-collimated electron beam, whose oscillations around the propagation axis are small. Previous work suggested this symmetry could be broken so as to increase $r_{\beta}$ by tailoring the laser wavefront\cite{mang09}. Here, we show through 3D simulations run with the particle-in-cell (PIC) code CALDER \cite{lefe03} that the use of a transverse density gradient distorts the wakefield, ultimately leading to an increase of the X-ray emission per charge unit from the betatron motion by a factor $>20$, when compared with simulations involving the acceleration of a similar charge to a similar energy in a plasma without the transverse gradient.

The paper is structured as follows. In Section II, we first focus the laser propagation in the transverse density gradient and detail the wakefield formation and the electron injection. In Section III, we then describe the motion of the trapped electrons, and show the increase of the transverse motion in the gradient. Section IV is dedicated to the study of the betatron radiation induced in the scheme, and Section V summarize our results.

\section{Laser propagation and Asymmetric wakefield generation}

In the proposed scheme, a 30~fs and 1.5~J laser with a wavelength $\lambda_0 = 0.8~\mu$m is focalized on a $W_0 = 13.5~\mu$m waist at the entrance of a 3.2~mm-long plasma characterized by a density $n_e = 3.5\times 10^{18}$~cm$^{-3}$. We use a time step $\Delta t = 0.248~\omega_0^{-1}$ and spatial steps $\Delta x = 0.25~c/\omega_0$ and $\Delta y = \Delta z = 4~c/\omega_0$ in the longitudinal and transverse directions respectively, with $c$ the speed of light and $\omega_0 = 2\pi c/\lambda_0$. In the simulations, the laser pulse is injected in a moving window at the transverse positions $z=0$ and $y=-10~\mu$m. It propagates with a small angle of incidence $\alpha$ from the $x$ axis in the $(x,y)$ plane. A flat density is used for $y<13~\mu$m, and a sharp density gradient rises from the transverse position $y=13~\mu$m. In the following simulation, $\alpha=3\degree$ and the density gradient is $0.42\times 10^{18}$~cm$^{-3}/\mu$m spanning on $50~\mu$m (density multiplied by 6 after $50~\mu$m), close to what can be experimentally achieved by inserting razor blades at the output of the gas nozzle \cite{schm10, buck13, guil15}. The laser pulse initially propagates in a homogeneous plasma, driving a classic ion cavity in its wake. When entering the gradient, the upper part of the laser pulse (i.e. at a higher $y$ position) propagates in a higher plasma density than its lower part. The wakefield period, which is roughly equal to the plasma wavelength $\lambda_p = 2\pi c/\omega_p$, with $\omega_p = \sqrt{n_ee^2/\varepsilon_0m_e}$ the plasma pulsation, where $e$ is the electron charge, $m_e$ the electron mass and $\varepsilon_0$ the vacuum permittivity, tends to vary transversally in the transverse density gradient. As a consequence, this creates an asymmetric wakefield [Fig. \ref{champlas}(a)], triggering an electron injection as will be developed later. Besides, the variation of the index of refraction in the gradient gradually deflects the laser pulse, which globally follows a parabolic trajectory given by a theoretical light ray interacting with the gradient, and is finally reflected [Fig. \ref{champlas}(d)]. However contrarily to the front of the laser pulse, its rear propagates in the ion cavity empty of electrons so it isn't influenced by the gradient, which leads to the tilted beam observed in Fig. \ref{champlas}(b). This in turn modifies the ponderomotive force and the wakefield creation, increasing the asymmetry. The rear part of the pulse is finally deflected after having reached the upper edge of the cavity and then shifts transversally [Fig. \ref{champlas}(e)]. This causes a transverse shift of the bubble, increasing the transverse motion of the electrons, which will then perform high amplitude betatron oscillations in the homogeneous plasma once the laser exits the gradient [Fig. \ref{champlas}(c)].

\begin{figure}[t]
\centering
\includegraphics[width=0.5\textwidth]{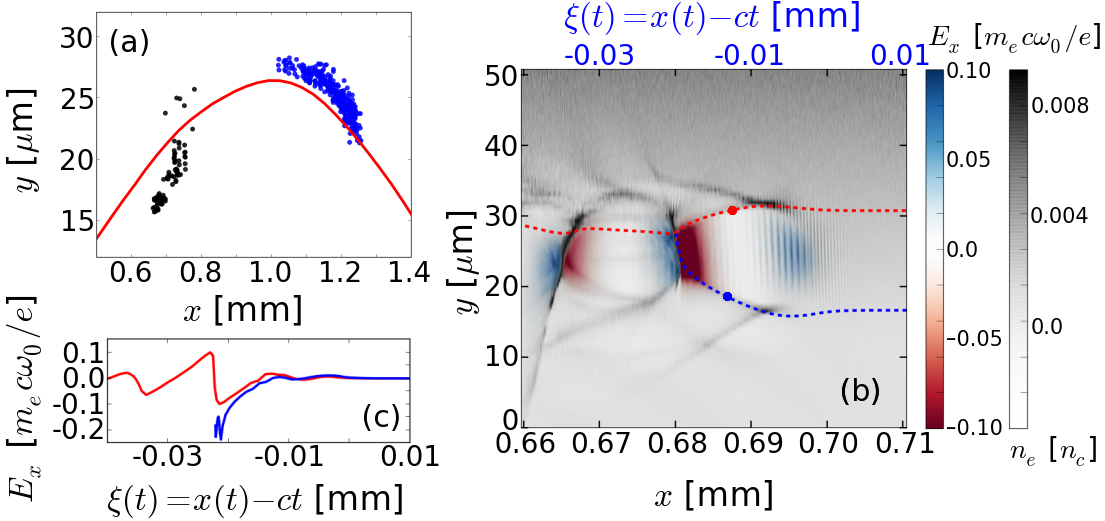}
\centering
\caption{Asymmetric injections in the transverse gradient. (a) Initial positions of the injected electrons in the ($x,y$) plane and trajectory of the laser beam barycenter (red solid line). The electrons injected during the first (resp. second) injection are plotted in black (resp. blue). (b) Spatial map of the plasma density (gray) and accelerating field $E_x$ (red-blue) after 0.69~mm of propagation. Trajectories of two particles (dashed lines) show the evolution of their position $y(t)$ as a function of $\xi(t) = x(t)-ct$. The dots represent their position after 0.69~mm of propagation. (c) Accelerating field $E_x$ experienced by the particles as a function of $\xi(t)$. The blue trajectory represents an injected particle while the particle following the red trajectory is not injected.}
\label{trajinj}
\end{figure}

The initial positions of the electrons injected during the propagation in the transverse gradient are shown in Fig. \ref{trajinj}(a), and suggest that two successive injections take place. The symmetry of the injection that usually occurs in LWFA is broken, similarly to the injection induced by asymmetric laser pulses \cite{cord13}. The first injection occurs shortly after the laser pulse enters the gradient at $x=0.7~$mm (black dots). These trapped electrons mainly originate from positions below the laser pulse. On the contrary, the second injection occurs between $x=1~$mm and $x=1.2$~mm (blue dots), just after the laser reaches its extremal position in $y$, and this second bunch of electrons comes from a position above the pulse.

These two bunches of electrons are injected by different mechanisms. We now focus on the first injected beam, which mainly contributes to the radiated energy as will be shown later. In Fig. \ref{trajinj}(b), the trajectories of a typical injected particle (blue line) and of a particle symmetrically situated on the other side of the laser (red line) are shown. Because of the asymmetric wakefield and the asymmetric ponderomotive force of the tilted laser pulse, the two particles do not have a similar trajectory. The injected particle is first ejected below the laser pulse by its ponderomotive force, and follows the lower edge of the cavity. It then crosses a zone with a strong accelerating field at the rear of the cavity [Fig. \ref{trajinj}(c)] and reaches a high enough energy to be injected [i.e. $\gamma>\gamma_s$, where $\gamma_s = \sqrt{2/3}(\omega_0/\omega_p)$ is the Lorentz factor associated with the wakefield]. On the contrary, the other particle ejected above the laser doesn't stay in the accelerating zone long enough to be injected. This first injection is then entirely provoked by the wakefield asymmetry, which elongates and amplifies the accelerating zone crossed by some of the electrons. The second injection [blue dots in Fig. \ref{trajinj}(a)] follows a more classical mechanism, essentially caused by the cavity elongation in the density downramp seen by the laser pulse when exiting the transverse gradient. Besides, self-injection is also observed further in the simulation after $x=2$~mm, but is of no importance as regards the X-ray emission, as we shall see later.

\begin{figure}[b]
\centering
\includegraphics[width=0.5\textwidth]{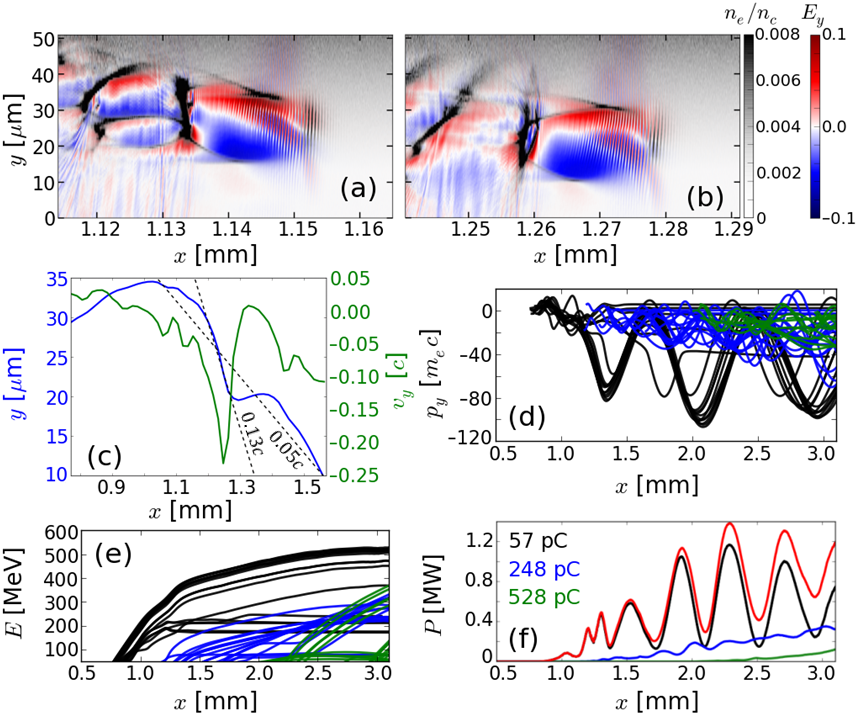}
\centering
\caption{Increase of the transverse motion induced by the transverse fields shifting. (a) (resp. (b)) Spatial map of the plasma density (gray) and transverse field $E_y$ (normalized to $m_ec\omega_0/e$, blue-red) after 1.16~mm (resp. 1.29~mm) of propagation. (c) Position $y$ where $E_y = 0$ at the rear of the accelerating cavity (blue line) and shifting speed $v_y$ of this position (green line). (d) (resp. (e)) Transverse momentum $p_y$ (resp. energy) of a few accelerated electrons issued from the first injection (black lines), the second injection (blue lines) and the third injection (green lines). 15 electrons are followed for each injection (not proportional to the injected charge). (f) Radiated power emitted by the electrons of each injection and total radiated power (red).}
\label{oscil}
\end{figure}

\section{Increase of the oscillation amplitude}

The wakefield asymmetry also causes an increase of the electron transverse motion. After the laser pulse has reached its extremal position in $y$, the rear of the laser interacts with the cavity edge and is finally deviated before the pulse leaves the gradient. As the back of the pulse reflects on a tilted cavity, it is reflected at a higher angle than the front of the pulse [as can be deduced from their respective position in Fig. \ref{champlas}(e)]. This induces a swift transverse displacement of the cavity and of the fields within [cf. Figs. \ref{oscil}(a) at $x = 1.16$~mm and \ref{oscil}(b) at $x = 1.29$~mm]. To quantify this transverse displacement, we represent the position where the transverse field $E_y$ is null near the rear of the cavity at the position of the injected electron bunch in Fig. \ref{oscil}(c) (blue line). Between $x=1.1~$mm and $x=1.6~$mm, the average speed of this position is $v_y=-0.05c$, in agreement with a $-3 \degree$ propagation angle. However, this speed climbs up to $v_y = -0.13c$ between $x=1.2~$mm and $x=1.3~$mm, due to the sweeping motion of the back of the cavity induced by the pulse back reflection. This transverse sweeping can give a significant transverse momentum to the already injected electrons. At $x\sim 1.1$~mm, electrons trapped during the first injection have already acquired an energy close to 200~MeV [Fig. \ref{oscil}(e)] and originally oscillate with a small transverse amplitude so that $|p_y|/m_ec<10$. But the transverse fields shifting is too fast and the electrons can't follow, as it occurs on a shorter spatial scale than a betatron motion period ($\lambda_{\beta}\sim500~\mu$m). The electron bunch will then be transversally accelerated between $x=1.2$~mm and $x=1.3$~mm. It acquires an average angle $\langle p_y \rangle/p_x\sim v_y/c=0.13$, which yields $\langle p_y \rangle/m_ec\sim 50$ for 200~MeV electrons. From an initial $\langle p_y \rangle/p_z\sim 0$ before $E_y$ shifts, $p_y$ finally oscillates with a $\sim50~m_ec$ amplitude around an average value $\sin(2\pi\alpha/360)p_x$, which is what we can observe in Fig. \ref{oscil}(d). For these electrons, the amplitude of the betatron motion reaches almost $10~\mu$m. This is much higher than the amplitude $\sim 1~\mu$m observed in a reference case with injection in a gradient with a flat transverse profile, and a triangular longitudinal shape with $70~\mu$m-long ramps and a maximal density $n_e = 5.2\times10^{18}$~cm$^{-3}$ (these properties are chosen so that the injected charges and final energies are similar to the transverse gradient case with $\alpha= 3\degree$). The electrons issued from later injections do not benefit from this effect as they aren't yet accelerated or injected when the swift transverse motion of the cavity occurs, and their transverse motions remain limited, with $|p_y|/m_ec<40$ for the second injection. As they also reach a lower maximal energy [Fig. \ref{oscil}(e)], they radiate a much weaker X-ray beam than the electrons trapped during the first injection [Fig. \ref{oscil}(f)], The first injected bunch accounts for the main part of the radiated power, leading to about 75\% of the total radiated energy ($3.6~\mu$J against $1.1~\mu$J for the electrons issued from the second injection), in spite of the lower charge $Q$ in this bunch (57~pC in the first injected bunch against 833~pC in total for the 3 bunches). Besides, the coherent betatron oscillations of the electrons lead to the oscillation observed in the radiated power.

\begin{figure}[t]
\centering
\includegraphics[width=0.5\textwidth]{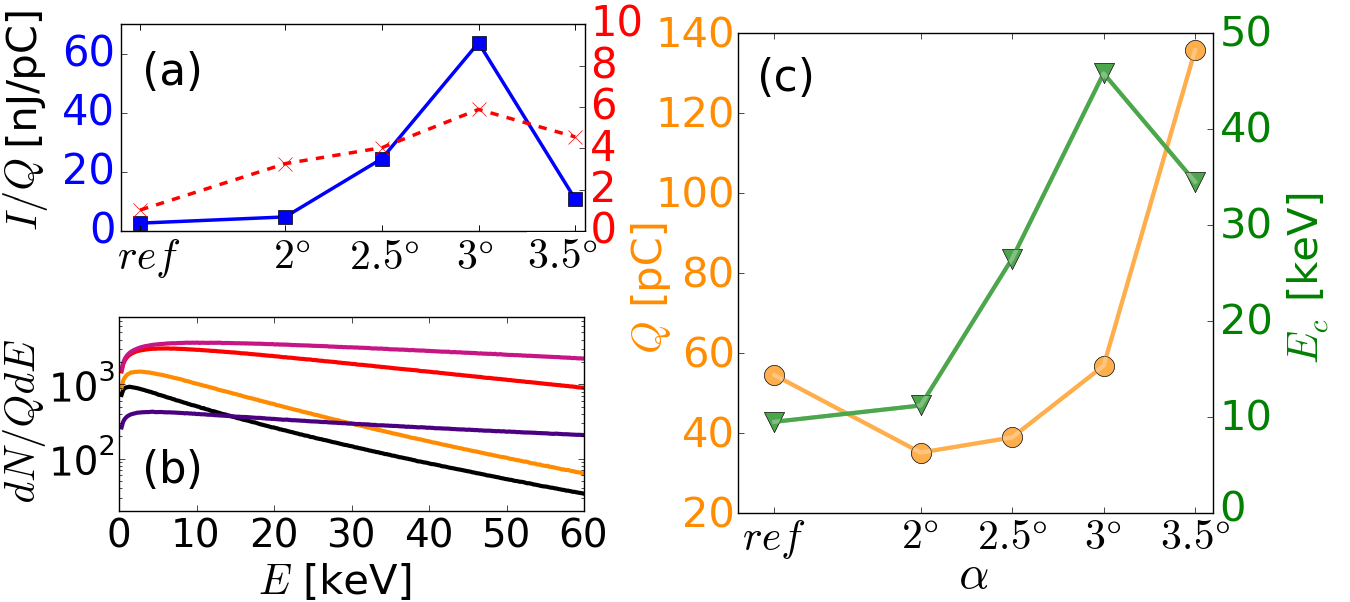}
\centering
\caption{Increase of the radiated energy in the transverse gradient scheme. (a) Radiated energy per unit of charge for the electrons trapped in the first-injected bunch (blue) and for the total injected charge (red). (b) Photon distribution per 0.1\%BW per unit of charge obtained after 3.1~mm of propagation for the electrons trapped during the first injection. Black lines: reference case (longitudinal gradient) and transverse gradient cases: $\alpha=2\degree$ (orange), $\alpha=2.5\degree$ (red), $\alpha=3\degree$ (pink) and $\alpha=3.5\degree$ (dark purple). (c) Charge trapped during the first injection (orange circles) and critical energy of the betatron source (green triangles) as a function of $\alpha$.}
\label{xray}
\end{figure}

\section{Betatron radiation}

As the laser beam doesn't propagate up to the upper limit of the $50~\mu$m-wide gradient, weaker gradients or higher angles could probably be used. We ran several simulations varying the angle $\alpha$ between $2\degree$ and $3.5 \degree$. Early injection in the gradient and increase of the transverse motion due to the wakefield transverse shifting are still observed in each case. Figure \ref{xray} shows the influence of varying $\alpha$ on the X-ray emitted by the electrons trapped during the first injection. Whatever the angle $\alpha$, we observe an increase of the X-ray signal emitted by these electrons when compared with the reference case -- for which we considered the front of the injected electron beam, which reaches a similar charge ($\sim50$~pC) and energy ($\sim450$~MeV). As the asymmetry effects increase for higher $\alpha$, so does the radiated energy per electron, which rises by a factor $\sim24$ for $\alpha=3\degree$ [Fig. \ref{xray}(a)] (63.7~nJ/pC compared with 2.7~nJ/pC in the reference case). When $\alpha>3\degree$, the laser pulse is however strongly perturbed while propagating in the gradient and part of the trapped charged which is accounted for in Fig. \ref{xray}(c) is then ejected transversally, reducing the number of electrons accelerated to high energy. This ejection mostly concerns the electrons with high wiggling amplitude, which filters out the most radiating electrons and reduces the radiated energy. This explains the decrease of the radiated energy for $\alpha=3.5\degree$. Photon distributions are represented in Fig. \ref{xray}(b), showing that the increase of the radiated energy corresponds particularly to an increase of the high energy photons emitted. These numerical photon distributions were fitted with theoretical synchrotron spectra \cite{four11b} of the form $dN/d\omega = S(\omega/\omega_c)$. We found $E_c = 9.5~$keV in the reference case and this critical energy increases drastically in the transverse gradient scheme [Fig. \ref{xray}(c)], reaching a maximal value of $45.8~$keV for $\alpha=3\degree$. Such a boost of the X-ray critical energy shall be of strong interest for medical applications, making the 10's of keV energy range accessible to a 1.5 Joule laser beam without aberrations \cite{ferr16}. Note however that whereas this scheme conserves the ultra-short duration of the betatron source, the amplified transverse motion will increase the source divergence in the direction of the gradient, as the divergence is directly proportionnal to $r_{\beta}$.

Considering the betatron source generated by the whole electron beam of 833~pC (729~pC in the reference case), this result is somehow mitigated by the weaker radiation emitted by the later injected charge, and the gain for the radiated energy per charge drops to a factor $>6$ between the case with $\alpha=3\degree$ and the reference case, as can be seen in Fig. \ref{xray}(a) (factor 3 for the critical energy). Nevertheless, this shall be further improved if the charge of the first injected beam in the transverse gradient is increased. Relying on any kind of injection scheme to inject an important charge before entering the transverse gradient could for example boost the full gain up to values close to 20. In that respect, ionization injection could be of particular interest in an experiment, similarly to what was done in a recent study exploiting the laser refraction on a shock front to increase the betatron oscillations \cite{yu18}. This could also be achieved by using other gradient profiles or by adapting the laser amplitude and waist when reaching the gradient.

\section{Conclusion}

In conclusion, we have demonstrated the validity of a new scheme improving the yield of a betatron source. It relies on the use of a sharp density gradient which has already been implemented in several experiments\cite{schm10, buck13, guil15}. The laser beam interaction with this gradient at a small angle can trigger the electron injection and boost the betatron motion amplitude thanks to a swift transverse shift of the wakefield in the gradient. X-ray emission per electrons has been shown to be significantly improved by a factor $>20$ compared with a reference case and the photon energies can be enhanced by a factor close to 5. Transverse gradients shall also be used with already injected electrons -- by self-injection or ionization-induced injection. Such electrons should still benefit from the transverse field shifting to increase their betatron motion. An other possibility to improve this scheme would be to change the gradient parameters. Different plasma densities or parabolic density profiles could for example be tested.

\section{Acknowledgments}

This work has been supported by Laserlab-Europe (EU-H2020 654148). We also acknowledge GENCI for awarding us access to TGCC/Curie under the grant No. 2016-057594 and the Knut and Alice Wallenberg Foundation.


%

\end{document}